\documentclass[extra,referee]{gji}
\usepackage{timet}
\usepackage{amsmath, enumitem}
\usepackage{amsfonts, bm}
\usepackage{upgreek}
\usepackage{graphicx}
\graphicspath{{Figures/}} 
\usepackage{setspace}
\usepackage{multirow}
\usepackage[T1]{fontenc}
\usepackage{fixltx2e}
\usepackage{xcolor}
\usepackage[skins,breakable]{tcolorbox}
\usepackage{caption}
\usepackage{gensymb}
\usepackage{hyperref}

\newtcolorbox{cvbox}[2][]{%
  blanker,
  width = 0.9\textwidth,
  after skip=8mm,
  title=#2,
  breakable,
  #1
}

\DeclareMathOperator*{\argmax}{arg\,max}
\DeclareMathOperator*{\argmin}{arg\,min}

\begin{document}

\begin{titlepage}
    \begin{center}
       \vspace*{-2.0cm} 
       
       
       \vspace{5cm}
       
       \huge
       \textbf{VIP - Variational Inversion Package with example implementations of Bayesian tomographic imaging}
       
       \vspace{4.0cm}
       \LARGE
       Xin Zhang$^{1,2}$ and Andrew Curtis$^{2}$ \\
       
       \vspace{1.0cm}
       \Large
       $^1$ School of Engineering and Technology, China University of Geosciences, Beijing, China \\
       $^2$ School of GeoSciences, University of Edinburgh, UK \\
       
       \vspace{1.0cm}
       \Large
       E-mail: \textit{x.zhang2@ed.ac.uk, andrew.curtis@ed.ac.uk}
       
       \vfill
       \vfill
    \end{center}
    
\end{titlepage}
\newpage

%
%
%
%
\begin{summary}
Bayesian inference has become an important tool to solve inverse problems and to quantify uncertainties in their solutions. Variational inference is a method that provides probabilistic, Bayesian solutions efficiently by using optimization. In this study we present a Python Variational Inversion Package (VIP), to solve inverse problems using variational inference methods. The package includes automatic differential variational inference (ADVI), Stein variational gradient descent (SVGD) and stochastic SVGD (sSVGD), and provides implementations of 2D travel time tomography and 2D full waveform inversion including test examples and solutions. Users can solve their own problems by supplying an appropriate forward function and a gradient calculation code. In addition, the package provides a scalable implementation which can be deployed easily on a desktop machine or using modern high performance computational facilities. The examples demonstrate that VIP is an efficient, scalable, extensible and user-friendly package, and can be used to solve a wide range of low or high dimensional inverse problems in practice.
\end{summary}

\section{Introduction}
In a variety of academic and practical applications that concern the Earth's subsurface we wish to find answers to specific scientific questions. In geosciences this is often achieved by imaging subsurface properties using data recorded on the surface, and by interpreting those images to address the question of interest. The subsurface is usually parameterized in some way, and a physical relationship is defined that predicts data and noise distributions that would be recorded for any particular set of model parameters, while the inverse relationship can not be determined uniquely. Once real data have been observed, the imaging problem is thus established as an inverse problem \citep{tarantola2005inverse}. 

Because of non-linearity in the physical relationship, insufficient data coverage and noise in the data, geophysical inverse problems almost always have non-unique solutions: many sets of parameter values can fit the data to within their uncertainty. It is therefore important to characterize the family of possible solutions (in other words, the solution uncertainty) in order to interpret the results with the correct level of confidence, and to provide well-justified and robust answers to the scientific questions \citep{arnold2018interrogation}.

Inverse problems are often solved by seeking an optimal set of parameter values that minimizes the difference or misfit between observed data and model-predicted data. Since most inverse problems have non-unique solutions, some form of regularization is often imposed on the parameters in order to make the computational solution unique \citep{aki1976determination,tarantola2005inverse,aster2018parameter}. Many codes and software have been developed using this class of methods \citep{rawlinson2005fmst, rucker2017pygimli, afanasiev2017salvus, xie2019specfem3d, wathelet2020geopsy}. However, since regularization is often chosen using ad-hoc criteria, these methods produce deliberately biased results, and valuable information can be concealed in the process \citep{zhdanov2002geophysical}. Moreover, no such optimization method can provide accurate estimates of uncertainty.

Bayesian inference solves inverse problems by updating a \textit{prior} probability density function (pdf) with new information contained in the data to produce a so-called \textit{posterior} pdf which describes the full state of information about the parameters post inversion \citep{tarantola2005inverse}. If we define the prior pdf as $p(\mathbf{m})$, the posterior pdf $p(\mathbf{m}|\mathbf{d}_{\mathrm{obs}})$ can be computed using Bayes' theorem:
\begin{equation}
	p(\mathbf{m}|\mathbf{d}_{\mathrm{obs}}) = \frac{p(\mathbf{d}_{\mathrm{obs}}|\mathbf{m})p(\mathbf{m})}{p(\mathbf{d}_{\mathrm{obs}})}
	\label{eq:Bayes}
\end{equation}      
where $p(\mathbf{d}_{\mathrm{obs}}|\mathbf{m})$ is the \textit{likelihood} function which describes the probability of observing the recorded data $\mathbf{d}_{\mathrm{obs}}$ if model parameters took the values in $\mathbf{m}$, and $p(\mathbf{d}_{\mathrm{obs}})$ is a normalization factor called the \textit{evidence}. This posterior pdf describes the full uncertainty in parameter values by combining the prior information and the uncertainty contained in the data.

Markov chain Monte Carlo (McMC) is one commonly-used method to solve Bayesian inference problems and has been used widely in many fields. The method constructs a set (chain) of successive samples that are distributed according to the posterior pdf by performing a structured random walk through parameter space \citep{brooks2011handbook}; thereafter, these samples can be used to estimate statistical information about parameters in the posterior pdf \citep{tarantola2005inverse} and to find answers to specific scientific questions \citep{arnold2018interrogation, siahkoohi2022deep, zhang2022interrogating, zhao2022interrogating, mckean2023separating}. The Metropolis-Hastings algorithm is one such method that originates from physics \citep{metropolis1949monte, hastings1970monte}, and has been applied to a range of geophysical applications \citep{mosegaard1995monte, malinverno2000monte, andersen2001bayesian, mosegaard2002monte, sambridge2002monte, ramirez2005stochastic, gallagher2009markov}. However, the algorithm becomes inefficient in high dimensional space because the chain can be attracted to individual maxima in the pdf due to its random-walk behavior. 

To improve the efficiency of McMC methods, a variety of more advanced methods have been introduced to geophysics, such as reversible-jump McMC \citep{green1995reversible, malinverno2002parsimonious, bodin2009seismic, galetti2015uncertainty, zhang20183}, Hamiltonian Monte Carlo \citep{duane1987hybrid, sen2017transdimensional, fichtner2018hamiltonian, gebraad2020bayesian}, Langevin Monte Carlo \citep{roberts1996exponential, siahkoohi2020faster}, stochastic Newton McMC \citep{martin2012stochastic, zhao2019gradient}, and parallel tempering \citep{hukushima1996exchange, dosso2012parallel, sambridge2013parallel}. Based on these studies a range of methods and codes have been developed to solve geophysical inverse problems using McMC \citep{bodin2009seismic, shen2012joint, hawkins2015geophysical, zhang20183, zunino2023hmclab}. Nevertheless, these papers mainly address 1D, 2D or sparsely-parametrised 3D spatial imaging problems; Bayesian solutions to large scale (high-dimensional) problems remain intractable because of their extremely high computational cost due to the curse of dimensionality \citep{curtis2001prior}.

In an attempt to improve the efficiency of Bayesian inference for certain types of problems, variational inference has been introduced to geophysics as an alternative to McMC. In variational inference one seeks a best approximation to the posterior pdf within a predefined family of (simplified) probability distributions by minimizing the difference between the approximating pdf and the posterior pdf \citep{bishop2006pattern, blei2017variational}. One commonly-used measure of the difference between the pdfs is the Kullback-Leibler (KL) divergence \citep{kullback1951information}. Variational inference therefore solves Bayesian inference problems by minimizing the KL divergence which is an optimization rather than a stochastic sampling problem. Consequently the method may be computationally more efficient and more scalable to high dimensionality in some classes of problems \citep{bishop2006pattern, zhang2018advances}. The method can also be applied to large datasets by dividing the data set into random minibatches and using stochastic and distributed optimization \citep{robbins1951stochastic,kubrusly1973stochastic}. By contrast, the same strategy cannot be used for McMC because it breaks the detailed balance condition required by most McMC methods \citep{o2004kendall}. In addition, variational inference methods can usually be parallelized at the individual sample level, whereas in McMC this cannot be achieved because of dependence between successive samples.

Variational inference has been applied to a range of geophysical applications. \cite{nawaz2018variational} used \textit{mean-field} variational inference to invert for subsurface geological facies distributions and petrophysical properties using seismic data, with further developments by \cite{nawaz2019rapid} and \cite{nawaz2020variational}. Although these methods are computationally extremely efficient, the mean-field approximation ignores correlations between parameters, and the methods of Nawaz and Curtis involved the development of bespoke mathematical derivations and implementations for each class of problem. While these developments resulted in speed of calculation, this approach restricts the method to a small range of problems for which correlations are not important and the derivations can be performed \citep{parisi1988statistical,bishop2006pattern, blei2017variational}. To extend variational inference to general inverse problems, \cite{kucukelbir2017automatic} used a Gaussian variational family to create a method called automatic differential variational inference (ADVI), which has been applied to travel time tomography \citep{zhang2020seismic} and earthquake slip inversion \citep{zhang2022bayesian}. By using a sequence of invertible and differential transforms (called normalizing flows), \cite{rezende2015variational} proposed normalizing flow variational inference in which those flows are designed so as to convert a simple initial distribution to an arbitrarily complex distribution that approximates the posterior pdf. In geophysics and related fields the method has been applied to travel time tomography \citep{zhao2022bayesian}, seismic imaging \citep{siahkoohi2020faster, siahkoohi2022wave}, seismic data interpolation \citep{kumar2021enabling}, transcranial ultrasound tomography \citep{orozco2023amortized} and cascading hazards estimation \citep{li2023disasternet}.

By using a set of samples of parameter values (called particles) to represent the density of an approximating pdf, \cite{liu2016stein} introduced a method called Stein variational gradent descent (SVGD) which iteratively updates those particle by minimizing the KL divergence so that the final particle density provides an approximation to the posterior pdf. SVGD has been demonstrated to be an efficient method in a range of geophysical applications, such as travel time tomography \citep{zhang2020seismic}, full waveform inversion (FWI) \citep{zhang2020variational, zhang2021bayesiana, lomas20233d, wang2023re}, earthquake source inversion \citep{smith2022hyposvi}, hydrogeological inversion \citep{ramgraber2021non}, post-stack seismic inversion \citep{izzatullah2023plug} and neural network based seismic tomography \citep{agata2023bayesian}. However the method becomes inefficient and inaccurate in high dimensional problems because of the finite number of particles and the practical limitation of computational cost \citep{ba2021understanding}. To reduce this issue, \cite{gallego2018stochastic} introduced the stochastic SVGD (sSVGD) method by combining SVGD and McMC: the efficiency of this method has recently been demonstrated when it was used to estimate the first Bayesian solution for a fully nonlinear, 3D FWI problems \citep{zhang20233}.

Despite these theoretical and practical advances, variational inference has not been widely used in geophysics. This is partly because the method is not easily accessible to non-specialists, and also because there is no common code framework to perform geophysical inversions using the method. In this study we therefore present a Python variational inversion package (VIP) which includes ADVI, SVGD and sSVGD, to make it more straightforward to solve geophysical inverse problems using variational inference methods. The package provides complete implementations of 2D travel time tomography and 2D full waveform inversion problems, including test results for users to check that their implementation is correct. Users can also solve other inverse problems by supplying their own forward functions and gradient calculation codes. In addition, to solve large inverse problems the package is designed in a scalable way such that it can be deployed on a desktop computer as well as in modern high performance computational (HPC) facilities.

In the following section we describe the concept of variational inference, and algorithmic details of ADVI, SVGD and sSVGD. In section 3 we provide an overview of the VIP package, and in section 4 we demonstrate VIP using examples of 2D travel time tomography and 2D full waveform inversion. We thus show that VIP is an efficient, scalable, extensible and user-friendly package that will enable users to solve geophysical inverse problems using variational methods. Making these methods more tractable for practitioners should allow them to be tested on a wide range of problems. 
 
\section{Theoretical background}
\subsection{Variational inference}
Within a predefined family of pdfs $Q=\{q(\mathbf{m})\}$, variational methods seek an optimal approximation $q^{*}(\mathbf{m})$ to the posterior probability distribution $p(\mathbf{m}|\mathbf{d}_{\mathrm{obs}})$ by minimizing the KL divergence between $q(\mathbf{m})$ and $p(\mathbf{m}|\mathbf{d}_{\mathrm{obs}})$:
\begin{equation}
	q^{*}(\mathbf{m}) = \argmin_{q \in Q} \mathrm{KL}[q(\mathbf{m})||p(\mathbf{m}|\mathbf{d}_{\mathrm{obs}})]
	\label{eq:argmin_KL} 
\end{equation}
where the KL divergence measures difference between two probability distributions:
\begin{equation}
\begin{aligned}
	\mathrm{KL}[q(\mathbf{m})||p(\mathbf{m}|\mathbf{d}_{\mathrm{obs}})] &= \mathrm{E}_{q}[\mathrm{log}q(\mathbf{m})] - \mathrm{E}_{q}[\mathrm{log}p(\mathbf{m}|\mathbf{d}_{\mathrm{obs}})] \\
	&= \mathrm{E}_{q}[\mathrm{log}q(\mathbf{m})] - \mathrm{E}_{q}[\mathrm{log}p(\mathbf{m},\mathbf{d}_{\mathrm{obs}})] + \mathrm{log}p(\mathbf{d}_{\mathrm{obs}})
\end{aligned}
\label{eq:KL}
\end{equation}
The KL divergence is non-negative and only equals zero when $q(\mathbf{m}) = p(\mathbf{m}|\mathbf{d}_{\mathrm{obs}})$ \citep{kullback1951information}. Because the evidence term $\mathrm{log}p(\mathbf{d}_{\mathrm{obs}})$ is computationally intractable, the KL divergence cannot be minimized directly. We therefore rearrange the above equation
\begin{equation}
\mathrm{log}p(\mathbf{d}_{\mathrm{obs}}) - \mathrm{KL}[q(\mathbf{m})||p(\mathbf{m}|\mathbf{d}_{\mathrm{obs}})]
 =  \mathrm{E}_{q}[\mathrm{log}p(\mathbf{m},\mathbf{d}_{\mathrm{obs}})] - \mathrm{E}_{q}[\mathrm{log}q(\mathbf{m})]
\label{eq:ELBO}
\end{equation}
Given that the KL divergence is non-negative, the left-hand side defines a lower bound of the evidence, called the evidence lower bound (ELBO):
\begin{equation}
\begin{aligned}
 	\mathrm{ELBO}[q] &= \mathrm{log}p(\mathbf{d}_{\mathrm{obs}}) - \mathrm{KL}[q(\mathbf{m})||p(\mathbf{m}|\mathbf{d}_{\mathrm{obs}})] \\
 	&= \mathrm{E}_{q}[\mathrm{log}p(\mathbf{m},\mathbf{d}_{\mathrm{obs}})] - \mathrm{E}_{q}[\mathrm{log}q(\mathbf{m})]
\end{aligned}
 	\label{eq:ELBO}
\end{equation}
Since the evidence $\mathrm{log}p(\mathbf{d}_{\mathrm{obs}})$ is a constant for a specific problem, minimizing the KL-divergence is equivalent to maximizing the ELBO. Variational inference in equation (\ref{eq:argmin_KL}) can therefore be expressed as:
\begin{equation}
	q^{*}(\mathbf{m}) = \argmax_{q \in Q} \mathrm{ELBO}[q(\mathbf{m})]
	\label{eq:argmax_elbo} 
\end{equation}

In variational inference, the choice of the variational family $Q$ is important because it determines both the accuracy of the approximation and the complexity of the optimization problem. Different choices of family may also allow different types of algorithm to be developed. In the VIP package we implement three different algorithms, ADVI, SVGD and sSVGD to solve inverse problems.
 
\subsection{Automatic differential variational inference (ADVI)}
ADVI uses the family of (transformed) Gaussians to solve variational inference problems \citep{kucukelbir2017automatic}. The transform arises because physical model parameters describe quantities that often have hard bounds, while Gaussian variables have infinite support. We therefore first transform the physical parameters into an unconstrained space using an invertible transform $T: \bm{\uptheta} = T(\mathbf{m})$. In this unconstrained space the joint distribution  $p(\mathbf{m},\mathbf{d}_{\mathrm{obs}})$ becomes:
\begin{equation}
	p(\bm{\uptheta},\mathbf{d}_{\mathrm{obs}}) = p(\mathbf{m},\mathbf{d}_{\mathrm{obs}})|\mathrm{det}\mathbf{J}_{T^{-1}}(\bm{\uptheta})|
	\label{eq:prob_transforma}
\end{equation}
where  $\mathbf{J}_{T^{-1}}(\bm{\uptheta})$ is the Jacobian matrix of the inverse of $T$. Define a Gaussian variational family
\begin{equation}
	q(\bm{\uptheta};\bm{\zeta})=\mathcal{N}(\bm{\uptheta}|\bm{\upmu},\bm{\Sigma})
	\label{eq:normal}
\end{equation}
where $\bm{\zeta}$ represents variational parameters, that is, the mean vector $\bm{\upmu}$ and the covariance matrix $\bm{\Sigma}$. To ensure that the covariance matrix $\bm{\Sigma}$ is positive semi-definite, we use a Cholesky factorization $\bm{\Sigma}=\mathbf{L}\mathbf{L}^{\mathrm{T}}$ where $\mathbf{L}$ is a lower triangular matrix, to reparameterize $\bm{\Sigma}$. 

With the above definition, the variational problem in equation (\ref{eq:argmax_elbo}) becomes:
\begin{equation}
	\begin{aligned}
		\bm{\zeta}^{*} &= \argmax_{\bm{\zeta}} \mathrm{ELBO}[q(\bm{\uptheta};\bm{\zeta})]\\
		&= \argmax_{\bm{\zeta}} \mathrm{E}_{q}[\mathrm{log}p\big(T^{-1}(\bm{\uptheta}),\mathbf{d}_{\mathrm{obs}}\big) + \mathrm{log}|det\mathbf{J}_{T^{-1}}(\bm{\uptheta})|] - \mathrm{E}_{q}[\mathrm{log}q(\bm{\uptheta};\bm{\zeta})] 
		\label{eq:argmaxELBO_advi}
	\end{aligned} 
\end{equation}  
As shown in \cite{kucukelbir2017automatic}, the gradients of the ELBO with respect to variational parameters $\bm{\upmu}$ and $\mathbf{L}$ can be calculated using:
\begin{equation}
	\nabla_{\bm{\upmu}}\mathrm{ELBO}
	=\mathrm{E}_{N(\bm{\upeta}|\mathbf{0},\mathbf{I})} 
	\big[
	\nabla_{\mathbf{m}}\mathrm{log}p(\mathbf{m},\mathbf{d}_{\mathrm{obs}}) 
	\nabla_{\bm{\uptheta}}T^{-1}(\bm{\uptheta})
	+ \nabla_{\bm{\uptheta}} \mathrm{log}|det\mathbf{J}_{T^{-1}}(\bm{\uptheta})| 
	\big]
	\label{eq:gradient_mu}
\end{equation}
\begin{equation}
	\nabla_{\mathbf{L}}\mathrm{ELBO}
	=\mathrm{E}_{N(\bm{\upeta}|\mathbf{0},\mathbf{I})} 
	\big[
	\big(
	\nabla_{\mathbf{m}}\mathrm{log}p(\mathbf{m},\mathbf{d}_{\mathrm{obs}}) 
	\nabla_{\bm{\uptheta}}T^{-1}(\bm{\uptheta})
	+ \nabla_{\bm{\uptheta}} \mathrm{log}|det\mathbf{J}_{T^{-1}}(\bm{\uptheta})| 
	\big) \bm{\upeta}^{\mathrm{T}}
	\big]
	+ (\mathbf{L}^{-1})^\mathrm{T}
	\label{eq:gradient_L}
\end{equation}
where $\bm{\upeta}$ is a random variable distributed according to the standard normal distribution $N(\bm{\upeta}|\mathbf{0},\mathbf{I})$. The expectations can be estimated using Monte Carlo (MC) integration, which in practice only requires a low number of samples because the optimization is performed over many iterations so that statistically the gradients will lead to convergence toward the correct solution \citep{kucukelbir2017automatic}. The variational problem in equation (\ref{eq:argmaxELBO_advi}) can therefore be solved by using gradient ascent methods. In the VIP package we implement four optimization algorithms: stochastic gradient descent (SGD), ADAGRAD \citep{duchi2011adaptive}, ADADELTA \citep{zeiler2012adadelta} and ADAM \citep{kingma2014adam}. The final approximation to the Bayesian solution can be obtained by transforming $q(\bm{\uptheta};\bm{\zeta}^{*})$ back to the original space. 

For transform $T$ we implement a commonly-used logarithmic transform \citep{stan2016stan, zhang2020seismic}
\begin{equation}
\begin{aligned}
 \theta_{i} &= T(m_{i}) = \mathrm{log}(m_{i}-a_{i}) - \mathrm{log}(b_{i}-m_{i}) \\
 m_{i} &= T^{-1}(\theta_{i}) = a_{i} + \frac{(b_{i}-a_{i})}{1+exp(-\theta_{i})}
\end{aligned}
 \label{eq:transform}
\end{equation} 
where $m_{i}$ and  $\theta_{i}$ represent the $i^{th}$ parameter in the original and transformed space respectively, and $a_{i}$ and $b_{i}$ are the lower and upper bound on $m_{i}$.

\subsection{Stein variational gradient descent (SVGD)}
SVGD is a variational method which uses a set of samples (called particles) to represent the approximating probability distribution. The particles are iteratively updated by minimizing the KL divergence so that the density of the final set of particles is distributed according to the posterior probability distribution. Define the set of particles as $\{\mathbf{m}_{i}\}$, then SVGD updates each particle using a smooth transform:
\begin{equation} 
T(\mathbf{m}_{i})=\mathbf{m}_{i} + \epsilon\bm{\upphi}(\mathbf{m}_{i})
\label{eq:svgd_transform}
\end{equation} 
where $\mathbf{m}_{i}$ is the $i^{th}$ particle, $\bm{\upphi}(\mathbf{m}_{i})$ is a smooth vector function which describes the perturbation direction, and $\epsilon$ is the magnitude of the perturbation. When $\epsilon$ is sufficiently small, the transform is invertible since the Jacobian of the transform is close to an identity matrix. Denote $q_{T}(\mathbf{m})$ as the transformed probability distribution of pdf $q(\mathbf{m})$. The gradient of the KL divergence between $q_{T}(\mathbf{m})$ and $p(\mathbf{m}|\mathbf{d}_{\mathrm{obs}})$ with respect to $\epsilon$ can be calculated as \citep{liu2016stein}:
\begin{equation}
\nabla_{\epsilon} \mathrm{KL}[q_{T}||p] \, |_{\epsilon=0} = - \mathrm{E}_{q} 
\left[ trace \left( \mathcal{A}_{p} \bm{\upphi} (\mathbf{m}) \right) \right] 
\label{eq:stein_gradient}
\end{equation} 
where $\mathcal{A}_{p}$ is the Stein operator defined as $\mathcal{A}_{p} \bm{\upphi}(\mathbf{m}) = \nabla_{\mathbf{m}} \mathrm{log} p(\mathbf{m}|\mathbf{d}_{\mathrm{obs}}) \bm{\upphi} (\mathbf{m})^{T} + \nabla_{ \mathbf{m} } \bm{\upphi} ( \mathbf{m} )$. This equation implies that one can obtain the steepest descent direction of the KL-divergence by maximizing the right-hand expectation, and consequently the KL divergence can be reduced by stepping a small distance in the reverse (negative) of that direction. Iteratively re-calculating equation (\ref{eq:stein_gradient}) and stepping in each revised direction locates a minimum in the KL divergence. 

The optimal direction $\bm{\upphi}^{*}$ that maximizes the expectation in equation (\ref{eq:stein_gradient}) can be calculated using \cite[see details in ][]{liu2016stein}:
\begin{equation}
	\bm{\upphi}^{*} \propto \mathrm{E}_{\{\mathbf{m'} \sim q\}} [\mathcal{A}_{p} k(\mathbf{m'},\mathbf{m})]
	\label{eq:phi_star}
\end{equation}
where $k(\mathbf{m'},\mathbf{m})$ is a kernel function which measures similarity between $\mathbf{m}$ and $\mathbf{m'}$ using their inner products. For example, a commonly-used kernel function is the radial basis function:
\begin{equation}
	k(\mathbf{m},\mathbf{m}') = \mathrm{exp} [- \frac{\Vert \mathbf{m}-\mathbf{m}' \Vert^{2}}{2h^{2}}]
	\label{eq:rbf}
\end{equation}
where $h$ is a scale factor that controls the magnitude of similarity between the two particles based on their distance apart. The expectation in equation (\ref{eq:phi_star}) can be estimated using the particles mean. We can therefore minimize the KL divergence by iteratively applying the transform $T(\mathbf{m})=\mathbf{m} + \epsilon\bm{\upphi}^{*}(\mathbf{m})$ to the set of initial particles $\{\mathbf{m}_{i}^{0}\}$:
\begin{equation}
	\begin{aligned}
		\bm{\upphi}^{*}_{l} (\mathbf{m}) &= \frac{1}{n} \sum_{j=1}^{n} \left[ k(\mathbf{m}_{j}^{l} , \mathbf{m}) \nabla_{\mathbf{m}_{j}^{l}} \mathrm{log} p(\mathbf{m}_{j}^{l}|\mathbf{d}_{\mathrm{obs}}) + \nabla_{\mathbf{m}_{j}^{l}} k(\mathbf{m}_{j}^{l}, \mathbf{m}) \right] \\
		\mathbf{m}_{i}^{l+1} &= \mathbf{m}_{i}^{l} + \epsilon^{l} \bm{\upphi}^{*}_{l} (\mathbf{m}_{i}^{l})
	\end{aligned}
	\label{eq:phi_mean}
\end{equation}
where $n$ is the number of particles and $l$ represents the $l^{th}$ iteration. For sufficiently small $\{\epsilon^{l}\}$, this process converges to the posterior distribution asymptotically as $n \to \infty$ \citep{liu2016stein}.

\subsection{Stochastic SVGD}
Stochastic SVGD (sSVGD) combines SVGD and McMC by adding a Gaussian noise term to the dynamics of SVGD. By doing this sSVGD becomes a McMC method with multiple interacting Markov chains, and since every set of particle values can be regarded as a sample of the posterior pdf, the method can generate many samples that are distributed according to the posterior pdf \citep{gallego2018stochastic}. sSVGD guarantees asymptotic convergence to the solution as the number of iterations tends to infinity, which standard SVGD with a finite number of particles cannot achieve. As a result sSVGD can produce more accurate results than the SVGD method, provided that the number of iterations is sufficient to remove effects of the distribution of samples near the start of the chain (the so-called burn-in period) \citep{gallego2018stochastic, zhang20233}.

For a random variable $\mathbf{z}$ denote the posterior distribution as $p(\mathbf{z})$; sSVGD simulates a stochastic differential equation (SDE) which converges to the distribution $p(\mathbf{z})$ \citep{ma2015complete}:
\begin{equation}
	d\mathbf{z} = \mathbf{f}(\mathbf{z})dt + \sqrt{2\mathbf{D}(\mathbf{z})}d\mathbf{W}(t)
	\label{eq:sde}
\end{equation}
with 
\begin{equation}
	\mathbf{f}(\mathbf{z}) = \left[\mathbf{D}(\mathbf{z}) + \mathbf{Q}(\mathbf{z}) \right]\nabla \mathrm{log}p(\mathbf{z}) + \Gamma(\mathbf{z})
	\label{eq:drift}
\end{equation}
where $\mathbf{W}(t)$ is a Wiener process, $\mathbf{D}(\mathbf{z})$ represents a positive semidefinite diffusion matrix, $\mathbf{Q}(\mathbf{z})$ is a skew-symmetric curl matrix, and $\Gamma_{i}(\mathbf{z}) = \sum_{j=1}^{d} \frac{\partial}{\partial\mathbf{z}_{j}}(\mathbf{D}_{ij}(\mathbf{z}) + \mathbf{Q}_{ij}(\mathbf{z}))$. To simulate this process, we discretize the above equation using the Euler-Maruyama discretization:
\begin{equation}
	\mathbf{z}_{t+1} = \mathbf{z}_{t} + \epsilon_{t} \left[ \left(\mathbf{D}\left(\mathbf{z}_{t}\right) + \mathbf{Q}(\mathbf{z}_{t})\right)\nabla \mathrm{log}p(\mathbf{z}_{t}) + \Gamma(\mathbf{z}_{t}) \right] + \mathcal{N}(\mathbf{0},2\epsilon_{t}\mathbf{D}(\mathbf{z}_{t}))
	\label{eq:discretized_sde}
\end{equation}
where $\mathcal{N}(\mathbf{0},2\epsilon_{t}\mathbf{D}(\mathbf{z}_{t}))$ represents a Gaussian distribution. The gradient $\nabla \mathrm{log}p(\mathbf{z}_{t})$ can be computed using the full data set, or using Uniformly randomly selected minibatch data subsets which results in a stochastic gradient approximation. In either case the above process converges to the posterior distribution asymptotically as $\epsilon_{t} \rightarrow 0$ and $t \rightarrow \infty$ \citep{ma2015complete}. Matrix $\mathbf{D(\mathbf{z})}$ and $\mathbf{Q}(\mathbf{z})$ can be adjusted to obtain faster convergence to the posterior distribution. For example, if set $\mathbf{D}=\mathbf{I}$ and $\mathbf{Q}=\mathbf{0}$, one obtains stochastic gradient Langevin dynamics \citep{welling2011bayesian}. If we construct an augmented space $\overline{\mathbf{z}}=(\mathbf{z},\mathbf{x})$ by concatenating a moment term $\mathbf{x}$ to the state space $\mathbf{z}$, and set $\mathbf{D}=\mathbf{0}$ and $\mathbf{Q}=\left( \begin{array}{cc} 
	\mathbf{0} & \mathbf{-I} \\ \mathbf{I} & \mathbf{0}
\end{array}\right)$ then the stochastic Hamiltonian Monte Carlo method can be derived \citep{chen2014stochastic}.

If we define an augmented space $\mathbf{z}=(\mathbf{m}_{1}, \mathbf{m}_{2}, ..., \mathbf{m}_{n})$ by concatenating the set of particles $\{\mathbf{m}_{i}\}$, equation (\ref{eq:discretized_sde}) can be used to generate samples from the posterior distribution $p(\mathbf{z})= \prod_{i=1}^{n} p(\mathbf{m}_{i}|\mathbf{d}_{\mathrm{obs}})$. Define a matrix $\mathbf{K}$ 
\begin{equation}
	\mathbf{K} = \frac{1}{n} \begin{bmatrix}
		k(\mathbf{m}_{1},\mathbf{m}_{1})\mathbf{I}_{d\times d} & \dots & k(\mathbf{m}_{1},\mathbf{m}_{n})\mathbf{I}_{d\times d} \\
		\vdots & \ddots & \vdots \\
		k(\mathbf{m}_{n},\mathbf{m}_{1})\mathbf{I}_{d\times d} & \dots & k(\mathbf{m}_{n},\mathbf{m}_{n})\mathbf{I}_{d\times d} 
	\end{bmatrix}
\label{eq:matrixK}
\end{equation}
where $k(\mathbf{m}_{i},\mathbf{m}_{j})$ is a kernel function defined in equation (\ref{eq:rbf}) and $\mathbf{I}_{d \times d}$ is an identity matrix. According to the definition of kernel functions, the matrix $\mathbf{K}$ is positive definite \citep{gallego2018stochastic}. By setting $\mathbf{Q}(\mathbf{z}_{t})=\mathbf{0}$ and $\mathbf{D}(\mathbf{z}_{t})=\mathbf{K}$, we obtain the so-called stochastic SVGD algorithm:
\begin{equation}
	\mathbf{z}_{t+1} = \mathbf{z}_{t} + \epsilon_{t} [\mathbf{K} \nabla \mathrm{log}p(\mathbf{z}_{t}) + \nabla \cdot \mathbf{K}]
	+ \mathcal{N}(\mathbf{0},2\epsilon_{t}\mathbf{K})
	\label{eq:stochastic_svgd}
\end{equation} 
Note that without the noise term $\mathcal{N}(\mathbf{0},2\epsilon_{t}\mathbf{K})$, the above equation becomes the standard SVGD method (compare equations \ref{eq:phi_mean} and \ref{eq:stochastic_svgd}). sSVGD is therefore a McMC method that uses the gradients from SVGD to produce successive samples. To generate samples from the distribution $\mathcal{N}(\mathbf{0},2\epsilon_{t}\mathbf{K})$ efficiently, we first define a matrix 
$\mathbf{D}_{\mathbf{K}}$
\begin{equation}
	\mathbf{D}_{\mathbf{K}} = \frac{1}{n} 
	\left[ \renewcommand\arraystretch{0.3}
	\begin{array}{ccc}
		\overline{\mathbf{K}} & & \\
		& \ddots & \\
		& & \overline{\mathbf{K}}
	\end{array}
	\right]
	\label{eq:matrixDK}
\end{equation}
where $\overline{\mathbf{K}}$ is a $n \times n$ matrix with $\overline{\mathbf{K}}_{ij} = k(\mathbf{m}_{i},\mathbf{m}_{j})$. The matrix $\mathbf{D}_{\mathbf{K}}$ can be constructed from $\mathbf{K}$ using $\mathbf{D}_{\mathbf{K}} = \mathbf{P}\mathbf{K}\mathbf{P}^{\mathrm{T}}$ where $\mathbf{P}$ is a permutation matrix

\begin{equation}
	\mathbf{P} = 
	\left[\arraycolsep=2pt \def\arraystretch{0.2}
	\begin{array}{clc|clc|c|clc}
		1 & & & & & & & & & \\
		& & & 1 & & & & & &  \\
		& & & & & & \ddots & & & \\
		& & & & & & & 1 & & \\
		\hline
		& 1 & & & & & & & & \\
		& & & & 1 & & & & &  \\
		& & & & & & \ddots & & & \\
		& & & & & & & & 1 & \\
		\hline
		& \ddots & & & \ddots & & \ddots & & \ddots & \\
		\hline
		& & 1 & & & & & & & \\
		& & & & & 1 & & & &  \\
		& & & & & & \ddots & & & \\
		& & & & & & & & & 1 \\
	\end{array}
    \right]
\end{equation}
The action of this permutation matrix on a vector $\mathbf{z}$ rearranges the order of the vector from the basis where the particles are listed sequentially to that where the first coordinates of all particles are listed, then the second, etc. With these definitions, a random sample $\bm{\eta}$ can be generated efficiently using 
\begin{equation}
\begin{aligned}
	\bm{\eta} &\sim \mathcal{N}(\mathbf{0},2\epsilon_{t}\mathbf{K}) \\
	&\sim \sqrt{2\epsilon_{t}} \mathbf{P}^{\mathrm{T}}\mathbf{P}\mathcal{N}(\mathbf{0},\mathbf{K}) \\
	&\sim \sqrt{2\epsilon_{t}} \mathbf{P}^{\mathrm{T}} \mathcal{N}(\mathbf{0},\mathbf{D}_{\mathbf{K}}) \\
	&\sim \sqrt{2\epsilon_{t}} \mathbf{P}^{\mathrm{T}} \mathbf{L}_{\mathbf{D}_\mathbf{K}}  \mathcal{N}(\mathbf{0},\mathbf{I})
\end{aligned}
\label{eq:noise_term}
\end{equation}
where $\mathbf{L}_{\mathbf{D}_\mathbf{K}}$ is the lower triangular Cholesky decomposition of matrix $\mathbf{D}_{\mathbf{K}}$. Taking account the fact that $\mathbf{D}_{\mathbf{K}}$ is a block-diagonal matrix, $\mathbf{L}_{\mathbf{D}_\mathbf{K}}$ can be computed easily as only the lower triangular Cholesky decomposition of matrix $\overline{\mathbf{K}}$ is required. In practice this calculation is computationally negligible because the number of particles $n$ is usually modest ($<$ 1000). One can now use equation (\ref{eq:stochastic_svgd}) to generate samples from the posterior distribution.

\section{Code overview}
The VIP package implements the suite of variational methods to solve geophysical inverse problems using the Python programming language. The package includes a set of specific forward and inverse problems such as 2D travel time tomography and 2D full waveform inversion, and also allows users to provide their own forward functions. In variational inference one needs to compute the gradient of the posterior pdf with respect to model parameters. We use the adjoint method to calculate the gradient in the case of seismic full waveform inversion \citep{lions1971optimal, tarantola1984inversion, tromp2005seismic, fichtner2006adjoint, plessix2006review}, and the ray tracing method in the case of travel time tomography \citep{rawlinson2004multiple}. For user-specified forward problems it is required that users implement their own function that computes gradients. 

The prior pdf is important in Bayesian inference as it provides information about model parameters independent of the data. The VIP package provides two commonly-used prior distributions: Uniform and Gaussian pdfs. To implement the Uniform distribution we employ two strategies. In the first strategy we impose hard constraints on model parameters, that is, for any parameter that assumes a value outside the distribution we reset the value to be the closest limit. Note that a similar strategy cannot be used in ADVI as the method assume a Gaussian variational family which cannot be defined in a constrained space. The second strategy involves using equation (\ref{eq:transform}) to transform model parameters into an unconstrained space and perform variational inversion in that space, which provides a more flexible way to employ a variety of variational families. In addition, users can provide their own prior distributions by implementing an appropriate pdf function (see details in the code documentation). 

Python is a popular high-level interpreted programming language which suffers from slow execution. We therefore implement time-consuming components of the code (e.g., the forward modelling functions) using Fortran and produce complied C extensions for these codes using the Cython framework \citep{behnel2010cython}. By doing this the code achieves C-like speeds. To further improve efficiency of the code, we use Dask to parallelize the forward computation at the sample (particle) level \citep{rocklin2015dask}. The package therefore provides an efficient, scalable and user-friendly implementation which can be deployed on a desktop as well as modern high performance computation facilities. Our aim is to implement a framework which can be used to solve various inverse problems, ranging from educational examples to complex, realistic studies.

\begin{figure}
	\includegraphics[width=1.0\linewidth]{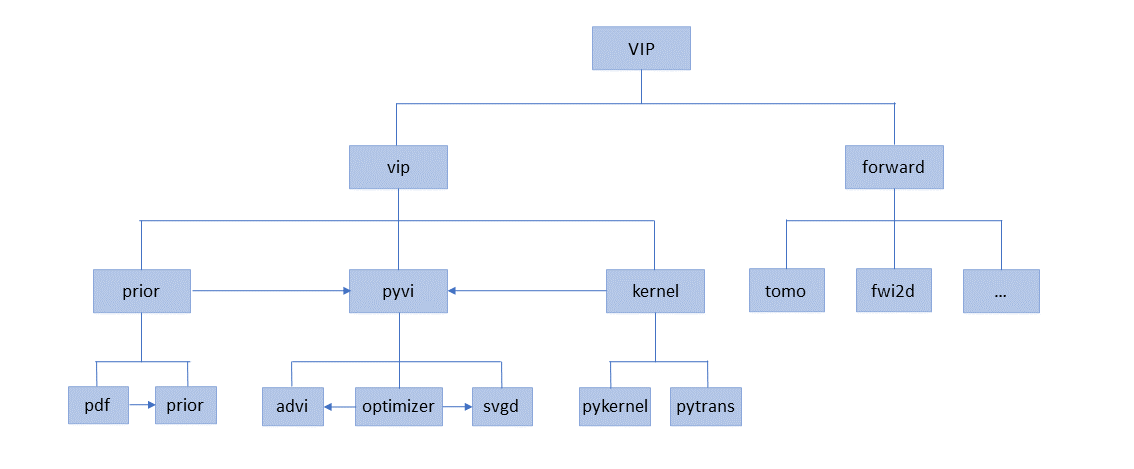}
	\caption{Code structure of VIP. Each rectangle represents a folder or file in the package. Users can implement their own forward functions similarly to the way this is implemented in examples \textit{tomo} and \textit{fwi2d}.}
	\label{fig:code_structure}
\end{figure}

Figure \ref{fig:code_structure} shows the structure of VIP. The inversion code (\textit{vip} in Figure \ref{fig:code_structure}) is implemented separately from forward modelling codes (\textit{forward} in Figure \ref{fig:code_structure}), and only requires an interface of forward functions that returns logarithmic posterior pdf values and gradients (details can be found in the code documentation and in two examples \textit{tomo} and \textit{fwi2d}). Thus, users can easily combine their own forward functions with the package. In the \textit{vip} code the prior distributions, kernel functions and variational algorithms are implemented in three different directories (\textit{prior}, \textit{kernel} and \textit{pyvi} in Figure \ref{fig:code_structure}) so that the code can be easily extended to other prior pdfs, kernel functions and variational methods. For example, users can implement their own prior pdfs by adding a proper pdf function in the \textit{pdf} code in the \textit{prior} directory. Note that both SVGD and sSVGD methods are implemented in the \textit{svgd} code.

\section{Applications}

\subsection{Travel time tomography}
\begin{figure}
\includegraphics[width=.9\linewidth]{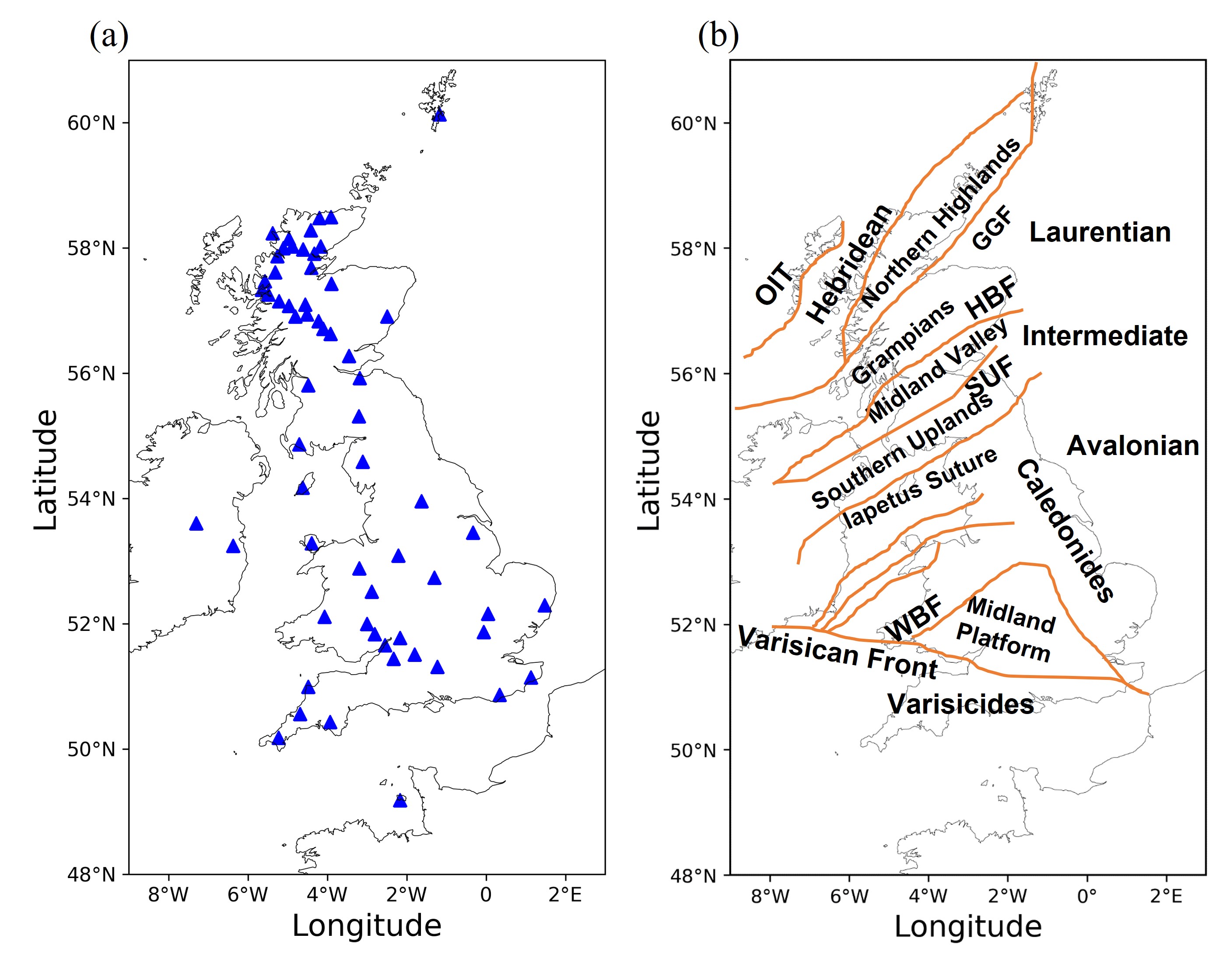}
\caption{\textbf{(a)} Locations of seismometers (blue triangles) around British Isles used in this study. \textbf{(b)} Terrane boundaries in the British Isles from \cite{galetti2017transdimensional}. Abbreviations are as follows: OIT, Outer Isles Thrust; GGF, Great Glen Fault; HBF, Highland Boundary Fault; SUF, Southern Uplands Fault; WBF, Welsh Borderland Fault System.}
\label{fig:uk_receivers}
\end{figure}

As a first example we use the VIP package to solve a 2D tomographic problem. Specifically, we create Love wave group velocity maps of the British Isles using ambient seismic noise data recorded by 61 seismometers (blue triangles in Figure \ref{fig:uk_receivers}a). The geological setting and the main terrain boundaries of the British Isles are shown in Figure \ref{fig:uk_receivers}b. The ambient noise data were recorded in 2001-2003, 2006-2007 and in 2010 using three different subarrays. The two horizontal components of the data (N and E) were first rotated to the transverse and radial directions, and the obtained transverse data were cross correlated to produce Love waves between different station pairs. Travel times of group velocity at different periods between different station pairs are then estimated from those love waves. Details of the data processing procedures can be found in \citep{galetti2017transdimensional}. In this study we use travel time measurements at 10 s period.

\begin{figure}
\includegraphics[width=1.\linewidth]{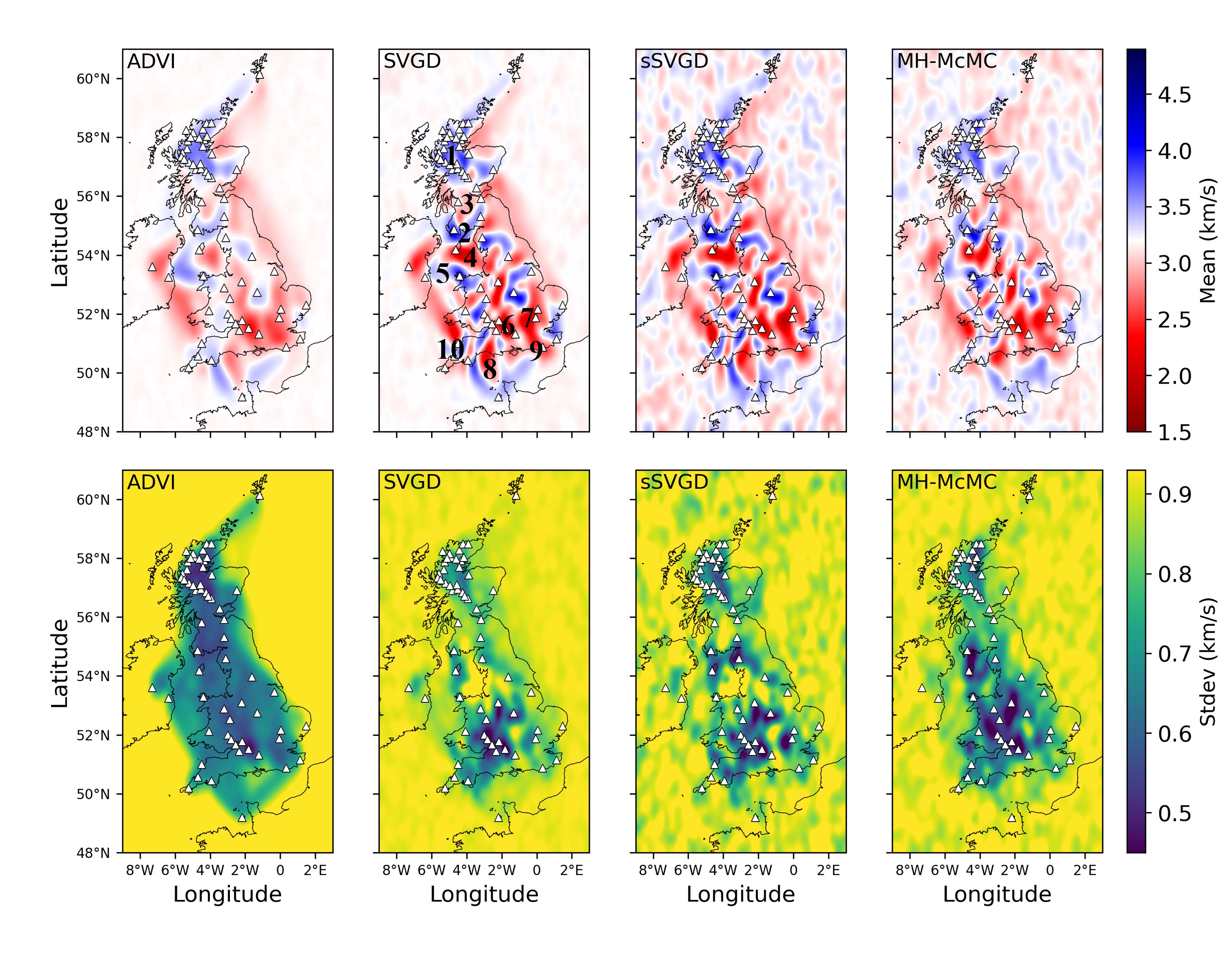}
\caption{Mean (top row) and standard deviation (bottom row) maps of group velocity at 10 s period obtained using ADVI, SVGD, sSVGD and MH-McMC respectively. White triangles denote locations of seismometers. Black numbers are referred to in the main text.}
\label{fig:uk_results}
\end{figure}

We parameterise the study region using a regular grid of 37 $\times$ 40 cells with a spacing of 0.33\degree in both longitude and latitude directions. The prior pdf for group velocity in each cell is set to be a Uniform distribution between 1.56 km/s to 4.8 km/s, of which the lower and upper bound were chosen to exceed the range of group velocities between all station pairs when assuming a great circle ray path \citep{zhao2022bayesian}. The likelihood function is chosen to be a Gaussian distribution to represent the data noise, which is estimated from independent travel time measurements by stacking randomly selected subsets of daily cross correlations \citep{galetti2017transdimensional}. In the inversion the predicted travel times are calculated using the fast marching method \citep{rawlinson2004multiple}.

We apply the above suite of methods to solve this tomographic problem, and compare the results with those obtained using the Metropolis-Hastings McMC (MH-McMC) method \citep{zhao2022bayesian}. The Uniform prior distribution is implemented using the second strategy that transforms variables into an unconstrained space in variational inversions. For ADVI, we started the method with a standard Gaussian distribution in the unconstrained space, and performed 10,000 iterations using the ADAM optimization algorithm \citep{kingma2014adam}. To visualize the results we generated 5,000 samples from the obtained Gaussian distribution and transformed them back to the original space to estimate posterior statistics. For SVGD, we generated 500 particles from the prior distribution and updated them using equation (\ref{eq:phi_mean}) for 3,000 iterations. The final particles are used to calculate the mean and standard deviation of the posterior distribution. For sSVGD, we started from 20 particles that are generated from the prior distribution, and updated them using equation (\ref{eq:stochastic_svgd}) for 6,000 iterations after an additional burn-in period of 2,000. To reduce the memory and storage cost, we only retained samples every fourth iteration after the burn-in period, which results in a total of 30,000 samples. 

Figure \ref{fig:uk_results} shows the mean and standard deviation maps obtained using the suite of variational methods, as well as those obtained using the MH-McMC algorithm \citep{zhao2022bayesian}. Overall the results obtained using different methods show similar mean structures which have a good agreement with the known geology and previous tomographic studies in the British Isles \citep{nicolson2012seismic, nicolson2014rayleigh, galetti2017transdimensional, zhao2022bayesian}. For example, in the Scottish highlands the mean maps clearly exhibits high velocities (annotation 1 in Figure \ref{fig:uk_results}) which are consistent with the distribution of Lewisian and Dalradian complexes in this area. Similarly high velocities associated with the accretionary complex of the Southern Uplands (annotation 2) are clearly visible around 4\textdegree W, 55\textdegree N following a SW-NE trend. Between the Highland Boundary Fault and the Southern Uplands Fault a similar tend of low velocity zone (annotation 3) is found in the Midland Valley. Low velocities are also observed in a number of sedimentary basins such as the East Irish Sea (4.5\textdegree W, 54\textdegree E - annotation 4), the Cheshire Basin (2.5\textdegree W, 52.5\textdegree E - annotation 6), the Anglian-London Basin (0\textdegree, 52\textdegree N - annotation 7), the Weald Basin (0\textdegree, 51\textdegree N - annotation 8) and the Wessex Basin (3\textdegree W, 50.5\textdegree N - annotation 9). By contrast, high velocities can be found in granitic intrusion regions, for example, in northwest Wales (around 4\textdegree W, 53\textdegree N - annotation 5) and Cornwall (around 4.5\textdegree W, 50.5\textdegree N - annotation 10). More detailed discussion and interpretation of the velocity structures can be found in \cite{galetti2017transdimensional}.

Among these results the mean map obtained using ADVI shows the smoothest structure, whereas other maps provide more detailed information. This has also been observed in previous studies \citep{zhang2020seismic, zhao2022bayesian} and is likely caused by the limitation of implicit Gaussian assumption made in ADVI. In offshore areas because few ray paths go through the open marine regions, the mean maps obtained using ADVI and SVGD show almost homogeneous velocity structure across these areas whose value is consistent with the mean of prior distribution. In comparison, the results obtained using sSVGD and MH-McMC have more heterogeneous structures, which probably indicates that the two methods have not converged sufficiently in these offshore areas: these areas are only loosely constrained by the data and hence have broad posterior uncertainties requiring more randomly generated samples in order to represent the posterior distribution accurately. Note that both sSVGD and MH-McMC involve random sampling of the posterior distribution, whereas samples obtained using SVGD are found deterministically by optimisation. As a result, SVGD produces smoother results \citep{zhang2021bayesiana, zhang20233}. 

Overall the standard deviation maps obtained using SVGD, sSVGD and MH-McMC show similar structures. For example, the results show lower uncertainties in the Scottish highlands and southern England because of dense arrays in those areas, whereas the offshore areas have higher uncertainties because few ray paths go through these regions. There is a high uncertainty loop around the low velocity anomaly in the Anglian-London Basin (annotation 7 in Figure \ref{fig:uk_results}), which has also been observed in previous studies \citep{galetti2015uncertainty,galetti2017transdimensional} and reflects uncertainty in the shape of the anomaly. By contrast, the standard deviation map obtained using ADVI shows different features. Although in the Scottish highlands the results still show lower uncertainty, the rest of the area within the receiver array has almost the same uncertainty level with little variation. In addition, in the West Irish Sea and the North Sea area between Northern Scotland and Shetland Islands the results show lower uncertainties which are not observed in the results obtained using other methods. This demonstrates that ADVI can produce biased results because of its underlying Gaussian assumption as found in previous studies \citep{zhang2020seismic}.  

Table \ref{tb:cost} compares the number of forward simulations required by each method to obtain these results, which provides a good metric of the computation cost as the forward simulation is the most computationally expensive component of each method. Note that the three variational methods require computation of derivatives of the posterior pdf with respect to model parameters which adds computational cost compared with the MH-McMC method. In this travel time tomography example the derivatives are calculated using ray paths, which are traced through the computed travel time field. This calculation requires a computation equivalent to approximately 0.08 forward simulations. We therefore compute the equivalent number of simulations by multiplying the number of simulations required by the three variational methods by 1.08, which are shown in third column in Table 1. 

The results indicate that ADVI is the most efficient method as it only requires 10,000 simulations, but we have demonstrated that the method probably produces biased results. SVGD demands the highest computational cost among the three variational methods, while sSVGD requires about 10 times fewer simulations than SVGD. This makes sSVGD a good choice for practical applications as noted in \cite{zhang20233}. Nevertheless, all three variational methods are significantly more efficient than the basic MH-McMC method implemented here as a bench-mark, which required 15 millions simulations in total with 10 independent parallel chains.

We note that the above comparison depends on subjective assessment of the point of convergence for each method, so the absolute number of simulations required by each method may not be entirely accurate (especially the number used for the MH-McMC algorithm). Nevertheless the comparison at least provides insights into the relative computational cost of each method. A more careful and thorough comparison between the same MH-McMC method and variational methods can be found in \cite{zhao2022bayesian} which again demonstrated that variational methods were computationally efficient. 

 \begin{table}
	\begin{center}
		\begin{minipage}{120mm}
			\caption{A comparison of computational cost for ADVI, SVGD, sSVGD and MH-McMC.} 	
			\begin{tabular}{c r r}
				\hline
				Method  & Number of simulations & Comparable number of simulations \\
				\hline
				ADVI  & 10,000 & 10,800 \\ 
				SVGD & 1500,000 & 1620,000 \\
				sSVGD & 160,000 & 172,800 \\
				MH-McMC & 15,000,000 & 15,000,000 \\
				\hline
			\end{tabular}
			\label{tb:cost}
		\end{minipage}
	\end{center}
\end{table}

\subsection{Full-waveform inversion}  

\begin{figure}
	\includegraphics[width=.9\linewidth]{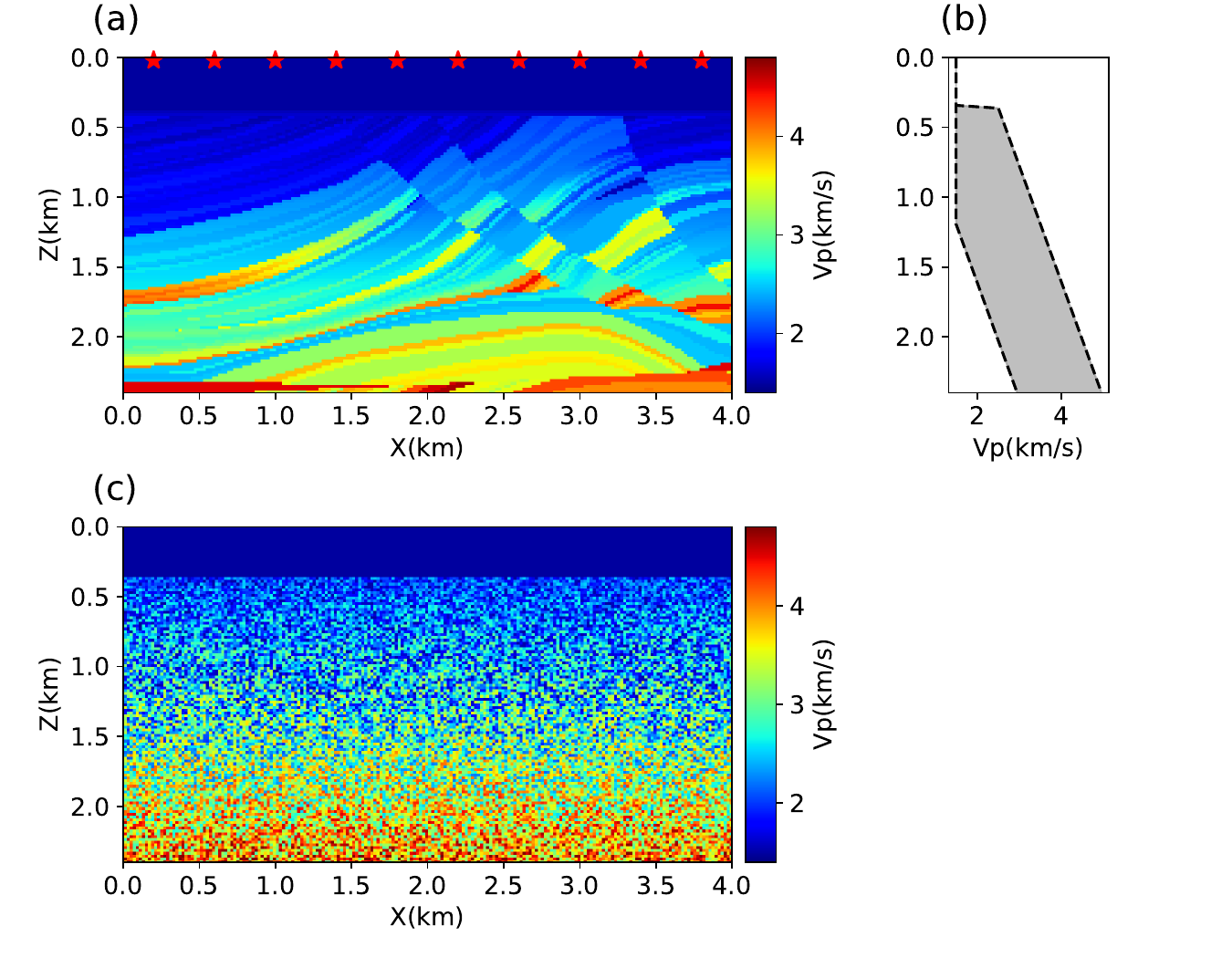}
	\caption{\textbf{(a)} The true model used in the full waveform inversion example. 10 sources are located at the depth of 20 m (red stars) and 200 receivers (not shown) are equally spaced at the depth of 360 m on the seabed. \textbf{(b)} The prior distribution of seismic velocity, which is set to be a Uniform distribution with an interval of 2 km/s at each depth. An additional lower bound of 1.5 km/s is also impose to the velocity to ensure that the rock velocity is higher than the velocity in water. \textbf{(c)} An example particle generated from the prior distribution.}
	\label{fig:fwi_prior}
\end{figure}

For the second example we use the VIP package to solve a 2D full waveform inversion problem. The true model is selected to be a part of the Marmousi model \cite[Figure \ref{fig:fwi_prior}a,][]{martin2006marmousi2}, and is discretized using a regular 120 $\times$ 200 grid with a spacing of 20 m. Ten sources are equally distributed at 20 m water depth (red stars in Figure \ref{fig:fwi_prior}), and 200 receivers are equally spaced at the depth of 360 m on the seabed across the horizontal extent of the model. We simulate the waveform data using a time-domain finite difference method with a Ricker wavelet of 10 Hz central frequency, and added Gaussian noise to the data whose standard deviation is set to be 2 percent of the median of the maximum amplitude of each seismic trace. The gradients of the logarithm posterior pdf with respect to velocity are calculated using the adjoint method \citep{tarantola1988theoretical, tromp2005seismic, fichtner2006adjoint, plessix2006review}. 

The prior distribution is set to be a Uniform distribution over an interval of 2 km/s at each depth (Figure \ref{fig:fwi_prior}b). To ensure that the rock velocity is higher than the velocity in the water, we imposed an extra lower bound of 1.5 km/s. For the likelihood function we use a Gaussian distribution to represent uncertainties on the waveform data:
\begin{equation}
	p(\mathbf{d}_{\mathrm{obs}}|\mathbf{m}) \propto \mathrm{exp}\left[-\frac{1}{2}\sum_{i}\left(\frac{d_{i}^{\mathrm{obs}}-d_{i}(\mathbf{m})}{\sigma_{i}}\right)^{2}\right]
	\label{eq:likelihood}
\end{equation}
where $i$ is the index of time samples, and $\sigma_{i}$ is the standard deviation of that sample.

We apply SVGD and sSVGD to solve this full waveform inversion problem as we have demonstrated that these methods provide more accurate results than ADVI. For SVGD we used 600 particles that are initially generated from the prior distribution (an example is shown in Figure \ref{fig:fwi_prior}c) and updated them using equation (\ref{eq:phi_mean}) for 600 iterations. The final particles are used to calculate statistics of the posterior distribution. For sSVGD we generated 20 particles from the prior distribution and updated them for 4,000 iterations after an additional burn-in period of 2,000. Similarly, to reduce the memory and storage cost we only retain samples from every tenth iterations, which results in a total of 8,000 samples. Those final samples are then used to compute statistics of the posterior distribution.

\begin{figure}
	\includegraphics[width=.9\linewidth]{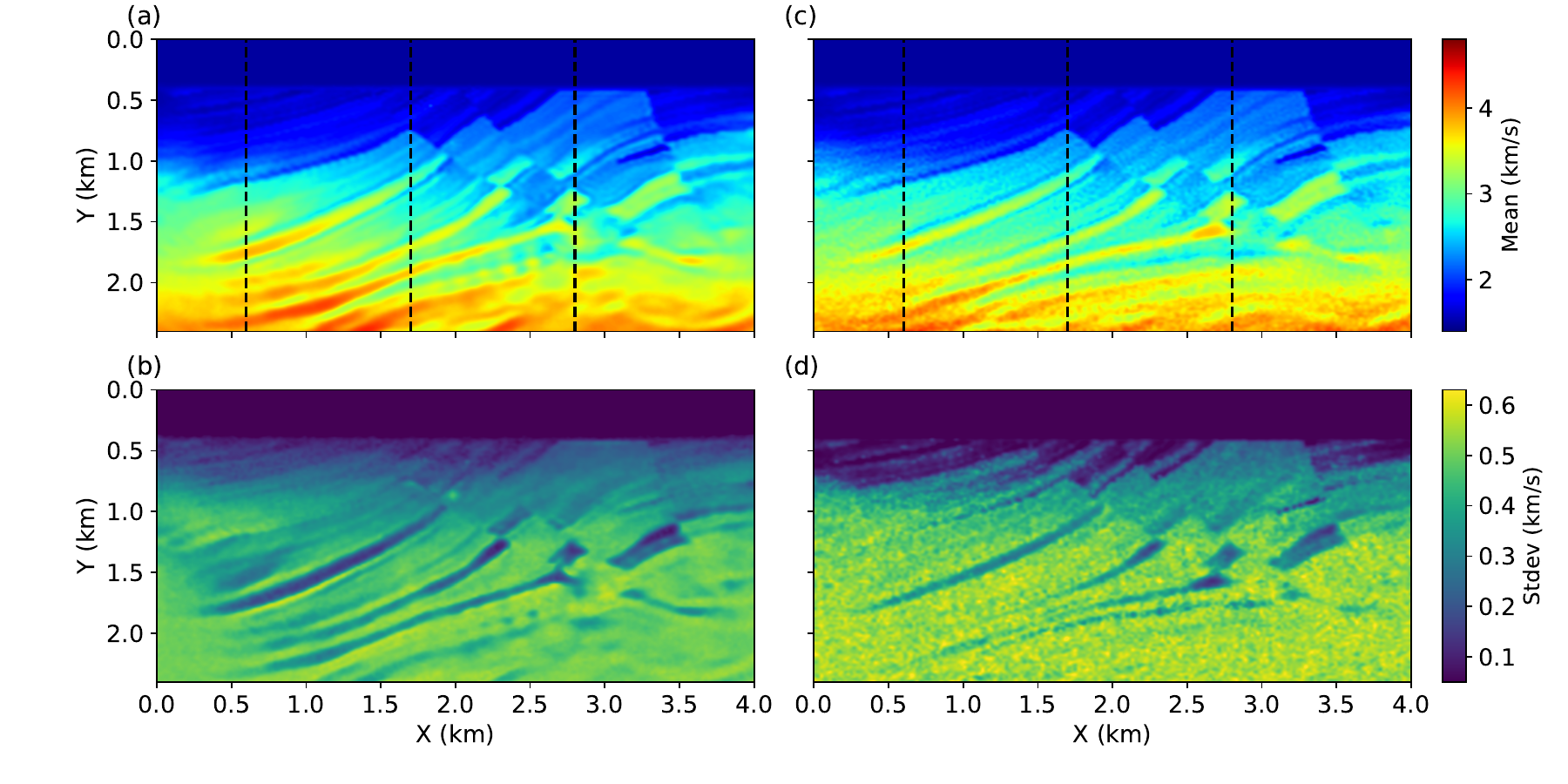}
	\caption{The mean (top row) and standard deviation (bottom row) obtained using SVGD (left panel) and sSVGD (right panel), respectively. Black dashed lines denote well log locations referred to in the main text.}
	\label{fig:fwi_mean_std}
\end{figure}

Figure \ref{fig:fwi_mean_std} shows the mean and standard deviation models obtained using SVGD and sSVGD. Overall the two methods produce similar results. For example, both mean models (Figure \ref{fig:fwi_mean_std}a and c) show similar structures to the true model, especially in the shallow part ($<$ 1.5 km). In the deep part ($>$ 1.5 km) and close to the sides, the mean models appear to be less similar to the true model because of lower resolution in those areas. However, the mean obtained using sSVGD is more similar to the true model than that obtained using SVGD. This reflects the fact that sSVGD can produce more accurate results than SVGD in high dimensional spaces, which has also been observed in other studies \citep{gallego2018stochastic, zhang20233}. Note that  similarly to the travel time tomography example above, the mean obtained using SVGD shows smoother structures than that obtained using sSVGD. This is likely because sSVGD is a McMC method which generates samples using stochastic sampling, whereas in SVGD particles are obtained deterministically using optimization. A similar phenomenon has also been observed in other studies when comparing results obtained using SVGD and sSVGD or McMC \citep{zhang2021bayesiana, zhang20233}.

\begin{figure}
	\includegraphics[width=.8\linewidth]{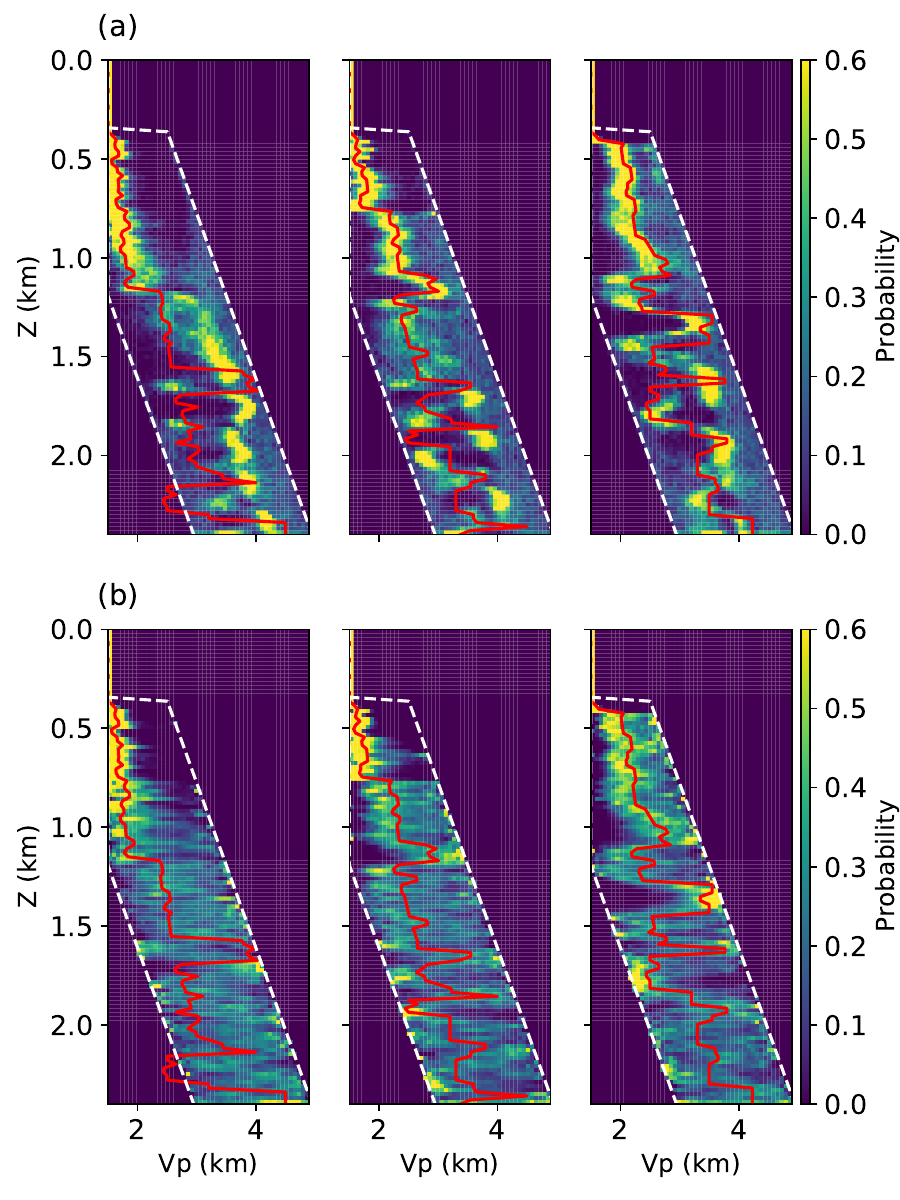}
	\caption{Marginal distributions at three well logs (black dashed lines in Figure \ref{fig:fwi_mean_std}) obtained using \textbf{(a)} SVGD and \textbf{(b)} sSVGD, respectively. Red lines show the true velocity profiles and white dashed lines show the lower and upper bound of the prior distribution.}
	\label{fig:fwi_marginals}
\end{figure}

Overall the standard deviation models show similar structural shapes to those in the mean model as has been observed in other studies \citep{gebraad2020bayesian, zhang2020variational, zhang2021bayesiana, zhang20233}. In the shallow part ($<$ 1.0 km) the results show lower uncertainties and in the deeper part the uncertainty is higher because of lower resolution. Those higher velocity anomalies in the deeper part are clearly associated with lower standard deviations, which likely reflects that those anomalies have large influences on the waveform data and hence have lower uncertainty. Similarly to the mean structures, the standard deviations obtained using SVGD show smoother structures than are obtained using sSVGD. In addition, the magnitude of the standard deviation obtained using SVGD is slightly lower than that obtained using sSVGD, which is likely because SVGD can underestimate uncertainties in high dimensional spaces due to the limited number of posterior samples produced \citep{ba2021understanding, zhang20233}. 

To further understand the results we show marginal distributions obtained using SVGD and sSVGD along three vertical profiles whose locations are denoted by dashed black lines in Figure \ref{fig:fwi_mean_std}. Overall the results show broader distributions in the deeper part ($>$ 1 km) than in the shallow part as we have observed in the standard deviation models. Furthermore, the distributions obtained using sSVGD are broader than those obtained using SVGD, which again demonstrates that SVGD can underestimate uncertainties. Note that in the results obtained using SVGD some true velocities lie outside the high probability area at large depths ($>$ 1.5 km), whereas those obtained using sSVGD generally include the true velocity in values with non-zero uncertainty. This shows that SVGD can produce biased results for high dimensional problems as noted in several studies \citep{ba2021understanding,zhang20233}.

\begin{table}
	\begin{center}
		\begin{minipage}{120mm}
			\caption{Computational cost required by SVGD and sSVGD for FWI.} 	
			\begin{tabular}{c c}
				\hline
				Method  & Number of simulations \\
				\hline 
				SVGD & 360,000 \\
				sSVGD & 120,000 \\
				\hline
			\end{tabular}
			\label{tb:cost2}
		\end{minipage}
	\end{center}
\end{table}

Similarly to the above section we measure the computational cost required by each method using the number of forward and adjoint simulations (Table \ref{tb:cost2}). Specifically, SVGD required 360,000 simulations to converge, while sSVGD used 120,000 simulations. This again demonstrates that sSVGD can be more computationally efficient than SVGD because sSVGD requires fewer particles yet generates many more samples. To give an overall idea of the computational cost, the above inversions required 49 hours for sSVGD using 40 AMD EPYC CPU cores, and 3 days for SVGD using 90 CPU cores.

\section{Discussion}
Although in the VIP package we only implemented 2D travel time tomography and 2D full waveform inversion, the code can easily be applied to larger scale problems by using modern high performance computation (HPC) facilities. For example, users can implement 3D full waveform inversion by providing a 3D forward and adjoint simulation code \cite[see more details in the code documentation, and an example in ][]{zhang20233}. In order to enable easy deployment on HPC facilities, the code provides a guide on how to parallelize the computation using the Sun Grid Engine queuing system. Other queuing systems can be implemented in a similar way.

Although we have demonstrated that sSVGD can generate more accurate results than SVGD in high dimensional problems and requires less computational cost in total, the method generally requires many more iterations. As a result, sSVGD may be less efficient than SVGD in wall clock time when a large number of CPU cores is available. This is why we implement SVGD in the VIP package as in practice it may be a better choice for low dimensional problems. 

ADVI may become inefficient in a high dimensional space because of the increased size of the covariance matrix. To enable applications in such cases, we also implement a diagonal covariance matrix, that is, a mean-field approximation \citep{kucukelbir2017automatic}. In SVGD and sSVGD besides the radial basis function kernel used in above examples, the package also implements diagonal matrix-valued kernel functions which are constructed by combining a positive definite diagonal matrix $\mathbf{Q}$ and the radial basis function \citep{wang2019stein, zhang2021bayesiana}. The elements of $\mathbf{Q}$ can be set as the inverse of the variance calculated across particles \citep{zhang2021bayesiana}.

To promote reproducibility and show how to use the code, we included several examples along with the code which can be used to reproduce those results obtained in the above section. We encourage interested readers to begin with these examples to familiarize themselves with the code. Finally, we note that VIP is actively being developed and expanded, and contributions from the community are welcome.
\section{Conclusion}  
VIP is a Python package which solves general inverse problems using variational inference methods, including automatic differential variational inference (ADVI), Stein variational gradient descent (SVGD) and stochastic SVGD (sSVGD). The package is designed to be easy enough for beginners to use, and efficient enough to solve complex inverse problems. In addition, VIP is implemented in a scalable way such that it can be deployed on a desktop as well as in high performance computation facilities. We demonstrated the package using two examples: 2D travel time tomography and 2D full waveform inversion. Users can also use the package to solve their own inverse problems by providing an appropriate forward modelling and gradient calculation code. We conclude that VIP can be used to solve a wide range of inverse problems in practice. The most recent release of the code can be downloaded from GitHub (\href{https://github.com/xin2zhang/VIP}{https://github.com/xin2zhang/VIP}).
\begin{acknowledgments}
The authors thank the Edinburgh Imaging Project sponsors (BP and Total) and National Natural Science Foundation of China (42204055) for supporting this research. This work has made use of the resources provided by the Edinburgh Compute and Data Facility (http://www.ecdf.ed.ac.uk/).
\end{acknowledgments}

\bibliographystyle{gji}
\bibliography{bibliography}

\appendix

\label{lastpage}

\end{document}